\documentclass[lettersize,journal]{IEEEtran}
\usepackage{amsmath,amsfonts}
\usepackage{algorithmic}
\usepackage{algorithm}
\usepackage{array}
\usepackage[caption=false,font=normalsize,labelfont=sf,textfont=sf]{subfig}
\usepackage{textcomp}
\usepackage{stfloats}
\usepackage{url}
\usepackage{verbatim}
\usepackage{graphicx}
\usepackage{cite}
\usepackage{epstopdf}
\usepackage{color,makecell}
\begin{document}

\title{Design and Performance Analysis of Index Modulation Empowered AFDM System}
\author{Jing Zhu, \textit{Graduate Student Member, IEEE}, Qu Luo, \textit{Graduate Student Member, IEEE}, Gaojie Chen, \textit{Senior Member, IEEE}, Pei Xiao, \textit{Senior Member, IEEE}, and Lixia Xiao, \textit{Member, IEEE}

\thanks{
J. Zhu, Q. Luo, G. Chen, and P. Xiao are with 5G and 6G Innovation centre, Institute for Communication Systems (ICS) of University of Surrey, Guildford, GU2 7XH, UK (e-mail: \{j.zhu, q.u.luo, gaojie.chen, p.xiao\}@surrey.ac.uk).

L. Xiao is with the Wuhan National Laboratory for Optoelectronics, Huazhong University of Science and Technology, Wuhan, 430074, China (e-mail: lixiaxiao@hust.edu.cn).
}

}

\vspace{-3em}

\maketitle

\begin{abstract}
In this letter, we incorporate index modulation (IM) into affine frequency division multiplexing (AFDM), called AFDM-IM, to enhance the bit error rate (BER) and  energy efficiency (EE) performance. In this scheme, the information bits are conveyed not only by $M$-ary constellation symbols, but also by the activation of the chirp subcarriers (SCs) indices, which are determined based on the incoming bit streams.
Then, two power allocation strategies, namely power reallocation (PR) strategy  and power saving (PS) strategy, are proposed to enhance BER and EE performance, respectively. Furthermore, the average bit error probability (ABEP) is theoretically analyzed. 
Simulation results demonstrate that the proposed AFDM-IM scheme achieves better BER performance than the conventional AFDM scheme.
\end{abstract}

\vspace{-1em}

\section{Introduction}
\IEEEPARstart{T}{he} beyond fifth generation (B5G) and sixth-generation (6G) wireless networks are expected to fulfill the need for ultra-reliable, high data rate and low-latency communications in high mobility scenarios.
In high mobility propagation environments, wireless channels can vary rapidly over time due to the Doppler effect, posing challenges for reliable communications \cite{6G_1,6G_2}.
As such, the commonly used  orthogonal frequency division multiplexing (OFDM) modulation may not be a viable solution because of the significant inter-Carrier interference (ICI) caused by Doppler spread \cite{6G_3}.

To tackle this issue, a new chirp-based multicarrier waveform, called affine frequency division multiplexing (AFDM), was proposed in \cite{AFDM_1}, which is a compelling alternative to OFDM in high-mobility systems. In AFDM, information symbols are multiplexed on a set of orthogonal chirps that are tuned to accommodate the characteristics of the doubly dispersive channel, allowing a comprehensive and sparse representation of the delay-Doppler (DD) domain via the discrete affine Fourier transform (DAFT) and the inverse DAFT (IDAFT). One of the distinctive advantages of the AFDM is its ability to attain full diversity over doubly dispersive channels, a notable contrast to existing chirp-based waveforms like orthogonal chirp division multiplexing (OCDM) in \cite{OCDM_1}. The AFDM input-output relation in matrix form was presented in \cite{AFDM_3}. Two low-complexity detectors by exploring the sparsity property of the effective channel matrix of AFDM were proposed in 
\cite{AFDM_2}.  Furthermore, AFDM was applied to multiple-input multiple-output (MIMO) communication systems in \cite{AFDM_4}, to meet the high data requirements of future B5G/6G high mobility scenarios.

Index modulation (IM) has garnered increasing interest in recent years, it enhances spectral efficiency (SE) by leveraging the antennas/subcarriers/time slots/channel states indices to transmit extra information \cite{IM_1}. IM-aided systems offer distinctive benefits, such as high energy efficiency (EE), low hardware complexity and flexible system structures, by introducing extra dimensions in contrast to conventional modulation systems. The concept of IM has been integrated with OFDM, called OFDM-IM, where the information is transmitted through both the $M$-ary modulated symbols as well as the subcarriers (SCs) indices \cite{OFDM_IM_1,OFDM_IM_2,OFDM_IM_3}.

To enhance BER and EE performance,  a fundamental investigation on the synergistic  amalgamation of AFDM and IM, referred to as AFDM-IM, is necessary for the evaluation of  the theoretical performance  and provision of  design guidelines. To the best of our knowledge, this work is the first of its kind to exploit the potential of AFDM with IM (AFDM-IM) to facilitate reliable transmission in high-mobility communication systems. This work makes the following novel contributions: 1) A new AFDM-IM scheme is proposed to improve the bit error rate (BER) and EE  performance compared to the classic AFDM. 2) Two power allocation strategies, namely power reallocation (PR) strategy and power saving (PS) strategy, are designed with the aim of enhancing the BER and EE performance, respectively. 3) The average bit error probability (ABEP) is analytically derived. Simulation results highlight the superiority of the proposed AFDM-IM over the classic AFDM and OFDM-IM systems{\footnote{\textit{Notations:} 
${\tbinom{n}{k}} $ and $\left\lfloor  \cdot  \right\rfloor $  refer to the binomial coefficient and the floor operations, respectively. ${(\cdot)^H}$ and ${(\cdot)^{ - 1}}$ denote Hermitian transpose and inverse, respectively. $\rm{diag}\left(  \cdot  \right)$ and $E\left(  \cdot  \right)$ stand for the diagonal and expectation operations, respectively.}}.

\begin{figure*}[t]
\centering
\includegraphics[width=0.90\textwidth]{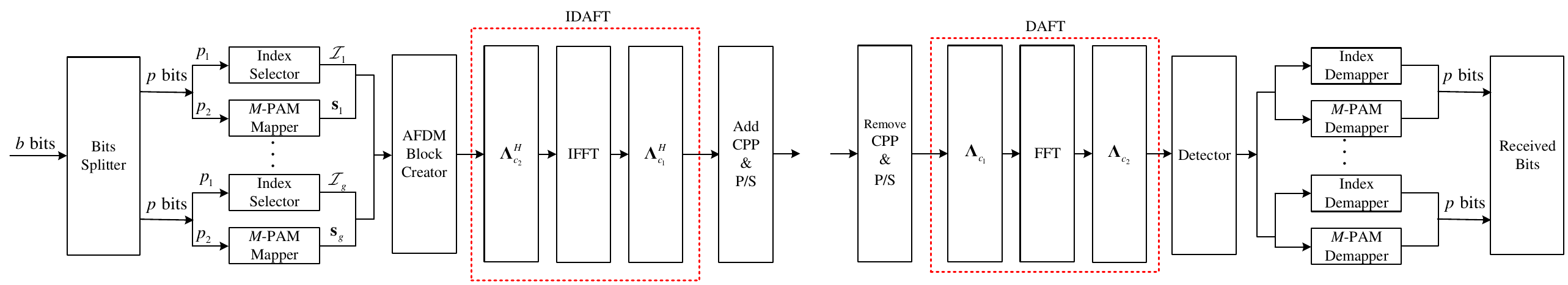}
\caption{Block diagram of the AFDM-IM system.}
\label{System_AFDM_IM}
\vspace{-1em}
\end{figure*}

\section{Proposed Affine Frequency Division Multiplexing with Index Modulation}
\subsection{System Model}
Consider an AFDM-IM scheme as shown in Fig. \ref{System_AFDM_IM}, a total of $b$ information bits enter the AFDM-IM transmitter for the transmission of each AFDM block. These $b$ bits are then equally divided into $G$ groups, with each group containing $p$ bits, i.e., $b=pG$. Each group of fixed $p$ bits are mapped to an AFDM subblock of length $N$, where $N=N_F/G$ and $N_F$ is the number of the AFDM SCs\footnote{Different from the OFDM system, which modulates the symbol with orthogonal sine wave SCs, the AFDM system uses a set of  orthogonal chirp SCs \cite{AFDM_3}.}. Unlike the conventional AFDM, the mapping operation of AFDM-IM involves modulated symbols mapping as well as the SCs indices mapping.
To be specific, for each subblock $g$, the incoming $p$ bits are first split into two parts. The first part of $p_1=\lfloor{{\log }_2}{ { {\tbinom{N}{K}} } }\rfloor$ bits are used for index selector to choose $K$ $(K<N)$ active indices out of $N$ available indices, where the selected indices are denoted as ${{\cal I}_g } = \{ {i_{g ,1}},...,{i_{g,k}},...,{i_{g,K}}\}$ with ${i_{g,k}} \in [1,...,N]$, $g=\{1,...,G\}$ and $k=\{1,...,K\}$. Note that the SCs index mapping process can be conducted using a look-up table for small $N$ and $K$. For example, Table I provides the case when $N=4$ and $K=2$.
The remaining ${p_2} = K{\log _2}(M)$ bits are modulated to $K$ $M$-ary phase shift keying (PSK) constellations, given by ${\bf{a}}_g=[{a_{g ,1}},...,{a_{g,k}},...,{a_{g,K}}]$ with ${a_{g,k}} \in {\cal A}$, $g=\{1,...,G\}$, $k=\{1,...,K\}$ and ${\cal A}$ is the PSK constellation set. Therefore, the transmitted signal vector of the $g$th subblock ${{\bf{x}}_g } \in {\mathbb{C}}^{N\times1}$ can be written as
\begin{equation}
{{\bf{x}}_g } \!= \!{[\underbrace {0,...,0}_{{i_{g ,1}} - 1},{a_{g ,1}},\underbrace {0,...,0}_{{i_{g ,k }} - {i_{g ,1}} - 1},{a_{g,k }},\underbrace {0,...,0}_{{i_{g ,K}} - {i_{g ,k }} - 1},{a_{g ,K}},\underbrace {0,...,0}_{N - {i_{g ,K}}}]^T}.
\end{equation}
After acknowledging both the index information and the signal constellations of all the subblocks, the AFDM block creator generates an $N_F \times 1$ discrete affine Fourier domain symbol as
\begin{equation}\label{AFDM_1}
{\bf{x}} = [{{\bf{x}}_1},...,{{\bf{x}}_g },...,{{\bf{x}}_G}].
\end{equation}
The $N_F$ point IDAFT is performed to map $\bf{x}$ to the time domain signal as
\begin{equation}\label{AFDM_2}
s[n] = \frac{1}{{\sqrt {N_F} }}\sum\limits_{m = 0}^{N_F - 1} {x[m]{e^{j2\pi ({c_1}{n^2} + {c_2}{m^2} + \frac{{nm}}{N_F})}}},
\end{equation}
where $c_1$ and $c_2$ are the DAFT parameters, and $n=0,...,N_F-1$.
Similarly to OFDM, the AFDM also requires a prefix for combating multi-path propagation.  Different from the OFDM, where a cyclic prefix (CP) is utilized, the AFDM employs a chirp-periodic prefix (CPP) due to the inherent periodicity of  DAFT. The CPP with a {length} of $L_{cp}$ is given by
\begin{equation}
s[n] = s[N_F + n]{e^{ - j2\pi {c_1}({{N_F}^2} + 2N_Fn)}},n =  - {L_{cp}},..., - 1.
\end{equation}

\begin{table}
\centering
\caption{The SCs index mapper for $N=4$ and $K=2$.}
{\begin{tabular}{|c|c|c|}\hline
Input bits & Indices & {AFDM-IM subblocks $({\bf{x}}_g)^T$} \\\hline
00 & $\{1,2\}$ &{$[a_{g,1} \ a_{g,2} \ 0 \ 0]$} \\\hline
01 & $\{2,3\}$ & {$[0 \ a_{g,1} \ a_{g,2} \ 0]$} \\\hline
10 & $\{3,4\}$ & {$[0 \ 0 \ a_{g,1} \ a_{g,2}]$}\\\hline
11 & $\{1,4\}$ & {$[a_{g,1} \ 0 \ 0 \ a_{g,2}]$}\\\hline
\end{tabular}}
\end{table}

Upon transmission over the channel and discarding CPP, the received samples can be written as
\begin{equation}\label{AFDM_3}
r[n] = {\sqrt{\rho}}\sum\limits_{l = 0}^\infty  {s[n - l]{g_n}[l]}  + w[n],
\end{equation}
where $\rho$ represents the average transmit power allocated to each active chirp SC. $w[n] \sim {\cal{CN}}(0,{N_0})$ is the additive Gaussian noise and
\begin{equation}
{g_n}[l] = \sum\limits_{l = 1}^P {{h_i}{e^{ - j\frac{{2\pi }}{{{N_F}}}{\varepsilon _i}n}}\delta (l - {l_i})}
\end{equation}
is the impulse response of channel at time $n$ and delay $l$, where $P \ge 1$ is the number of path, $\delta ( \cdot )$ is the Dirac delta function, $h_i$ is the complex channel gain of the $i$th path, obeying the distribution of ${\cal{CN}}(0,1/P)$, $l_i \in [0,l_{\rm{max}}]$ is the integer delay associated with the $i$th path with the maximum delay $l_{\rm{max}}$, and $
{\varepsilon _i} = {\alpha _i} + {\beta _i}\in [ - {\varepsilon _{\max }},{\varepsilon _{\max }}]$ is the Doppler shift normalized with respect to the SC spacing $\Delta f$, where ${\alpha _i}\in [ - {\alpha _{\max }},{\alpha _{\max }}]$ denotes its integer part and ${\beta _i}\in ( - \frac{1}{2},\frac{1}{2}]$ is its fractional part \cite{AFDM_3}.

Performing $N_F$ point DAFT on the received samples $r[n]$, we can obtain the output symbols in  the discrete affine Fourier domain  as
\begin{equation}
{
y[m] = \sqrt \frac{{\rho}}{N_F}\sum\limits_{n = 0}^{N_F - 1} {r[n]{e^{ - j2\pi ({c_1}{n^2} + {c_2}{m^2} + \frac{{nm}}{N_F})}}},}
\end{equation}
where $m=0,...,N_F-1$.

\vspace{-1em}
\subsection{Input-Output Relation in Matrix Form}
According to \eqref{AFDM_1}-\eqref{AFDM_2}, the transmitted signal in time domain $\bf{s}$ can be expressed as
\begin{equation}
\mathbf s = {\bf{\Lambda }}_{c_1}^{H} \mathbf F^{H} {\bf{\Lambda }}_{c_2}^{H}  \mathbf x = \mathbf A^{H} \mathbf x,
\end{equation}
where ${\bf{A}}={{\bf{\Lambda }}_{{c_2}}}{\bf{F}}{{\bf{\Lambda }}_{{c_1}}}$ is the DAFT matrix and $\mathbf A^{H}$ denotes the IDAFT matrix. $\bf{F}$ is the discrete Fourier transform (DFT) matrix with entries ${e^{ - j2\pi mn/N_F}}/\sqrt {N_F} $ and ${{\bf{\Lambda }}_{c_i}} = {\rm{diag}}({e^{ - j2\pi {c_i}{n^2}}},n = 0,...,N_F - 1)$.

At the receiver, the received signal is written as
\begin{equation}
{\bf{r}} = {\sqrt{\rho}}{\bf{Hs}} + {\bf{w}},
\end{equation}
where ${\bf{w}} \sim {\cal{CN}}(0,{N_0}{\bf{I}})$, ${\bf{H}} = \sum\nolimits_{i = 1}^P {{h_i}{{{\bf{\Gamma }}_{{\rm{CP}}{{\rm{P}}_i}}}}{{\bf{\Delta }}_{{\varepsilon_i}}}{{\bf{\Pi }}^{{l_i}}}}$, and ${\bf{\Pi }}$ is the forward cyclic-shift matrix.
${{\bf{\Delta }}_{{\varepsilon _i}}} = {\rm{diag}}({e^{ - j\frac{{2\pi }}{{{N_F}}}{\varepsilon _i}n}},n = 0,...,{N_F} - 1)$ and ${{{\bf{\Gamma }}_{{\rm{CP}}{{\rm{P}}_i}}}}$ is a $N_F \times N_F$ diagonal matrix
\begin{equation}
{{\bf{\Gamma }}_{{\rm{CP}}{{\rm{P}}_i}}} = {\rm{diag}}\left( {\left\{ {\begin{array}{*{20}{l}}
{{e^{ - j2\pi {c_1}({{N_F}^2} - 2{N_F}({l_i} - n))}},}&{n < {l_i}}\\
{1,}&{n \ge {l_i}}
\end{array}} \right.} \right),
\end{equation}
where $n=0,...,N_F-1$.


After performing $N_F$ point DAFT, the received signal in discrete affine Fourier domain is given by
\begin{equation}\label{AFDM_4}
\begin{aligned}
{\bf{y}} = {\bf{Ar}} &= {\sqrt{\rho}}\sum\limits_{i = 1}^P {{h_i}} \underbrace{{\bf{A}}{{\bf{\Gamma }}_{{\rm{CP}}{{\rm{P}}_i}}}{{\bf{\Delta }}_{{\varepsilon_i}}}{{\bf{\Pi }}^{{l_i}}}{{\bf{A}}^H}}_{\mathbf H_i}{\bf{x}} + {\bf{Aw}}\\
 &= {\sqrt{\rho}}{{\bf{H}}_{{\rm{eff}}}}{\bf{x}} + {\bf{\tilde w}},
\end{aligned}
\end{equation}
where ${{\bf{H}}_{{\rm{eff}}}} = \sum\limits_{i = 1}^P {{h_i}} \mathbf H_i$ , ${\bf{H}}_i =  {{\bf{A}}{{\bf{\Gamma }}_{{\rm{CP}}{{\rm{P}}_i}}}{{\bf{\Delta }}_{{\varepsilon_i}}}{{\bf{\Pi }}^{{l_i}}}{{\bf{A}}^H}}$ is the $i$th path channel in discrete affine Fourier domain,   and ${\bf{\tilde w}} = {\bf{Aw}}$.

\vspace{-1em}
\subsection{Power Allocation Strategy}
Assumed that the total transmit power $E_T$ is equally allocated to each SC of an AFDM block, so that the average power per SC in the conventional AFDM scheme is equal to $\rho  = \frac{{{E_T}}}{{{N_F}}}$. However, in AFDM-IM, only a subset of SCs are activated to convey information bits, addressing the allocation of power initially designated for inactive SCs is a pivotal problem in enhancing system performance.
To solve this issue, we propose two strategies, called power reallocation (PR) and power saving (PS), respectively.

\textit{Power Reallocation Strategy:} In the PR strategy,  the power initially allocated to the inactive SCs is equally redistributed among the active SCs. As a result, the power allocated to each active SC is ${\rho _{{\rm{PR}}}} = \frac{{N{E_T}}}{{K{N_F}}} = \frac{{N\rho }}{K}$. It is clear that the power allocated to each active SC is increased compared to the conventional AFDM, offering the BER performance enhancement, which will be verified in the simulation results.

\textit{Power Saving Strategy:} In the PS strategy, the power initially allocated to the inactive SCs is completely compressed. Hence, the power allocated to each active SC is still $\rho  = \frac{{{E_T}}}{{{N_F}}}$, but the total transmit power is reduced to $\frac{{{GKE_T}}}{{{N_F}}} $, resulting in better EE since $GK<N_F$.

\vspace{-1em}
\subsection{Signal Detection for AFDM-IM system}
\textit{1) ML detector:} The optimal ML detector for the proposed AFDM-IM system can be formulated as
\begin{equation}
{{\bf{\hat x}}_{ML}} = \mathop {\arg \min }\limits_{{\bf{x}}\in \mathbb{X}} \left\| {{\bf{y}} - {\sqrt{\rho}} {{\bf{H}}_{{\rm{eff}}}}{\bf{x}}} \right\|_F^2,
\end{equation}
where $\mathbb{X}$ is the set of the AFDM-IM transmitted vectors with the size of $2^b$. 

\textit{2) MMSE detector:} The MMSE detector is formulated as
\begin{equation}
{{\bf{\hat x}}_{MMSE}} = {({\bf{H}}_{{\rm{eff}}}^H{{\bf{H}}_{{\rm{eff}}}} + {N_0}{{\bf{I}}_{N_F}})^{ - 1}}{\bf{H}}_{{\rm{eff}}}^H{\bf{y}}.
\end{equation}
After obtaining the estimated signal ${{\bf{\hat x}}_{MMSE}}$, the bit information is detected by AFDM-IM demodulation.

\vspace{-1em}
\subsection{AFDM-IM Parameters}
The performance of AFDM-IM systems is highly sensitive to the selection of parameters $c_1$ and $c_2$. In (\ref{AFDM_4}), the element at the $p$th column and $q$th row of   the $i$th path  channel is defined as  \cite{AFDM_3}
 \begin{equation}
 \label{Hp}
 \begin{aligned}
 &  {{\mathbf{H}}_i}[p,q] = \\
 &\underbrace{e^{j\frac{{2\pi }}{N_F}\left( {N_F{c_1}l_i^2 - q{l_i} + N_F{c_2}\left( {{q^2} - {p^2}} \right)} \right)}}_{\eta(l_p,p,q) } \underbrace{ \frac{e^{-j2\pi (p-q+\text{Ind}_i + \beta_i)-1}}{e^{\frac{-j2\pi}{N_F} (p-q+\text{Ind}_i + \beta_i)-1}}}_{\gamma(l_i,\varepsilon_i,p,q)},
 \end{aligned}
 \end{equation}
  where $ \text{Ind}_i= (\alpha_i + 2N_F c_1 l_i)_{N_F}$.  For each path, $\eta(l_i,p,q) $ has unite energy,  and $\gamma(l_i,\varepsilon_i,p,q)$ achieves the peak energy at $q = (p+\text{Ind}_i)_{N_F}$ and decreases as $q$ moves always from  $(p+\text{Ind}_i)_{N_F}$.  In this paper,  $\gamma(l_i,\varepsilon_i,p,q)$ is considered to be non-zero only for $q$ moves  $k_{\varepsilon}$ always from  $(q+\text{Ind}_i)_{N_F}$.  In AFDM-IM,  it is essential to determine  $c_1$ and $c_2$ as such the impulse response in the DAFT domain constitutes a comprehensive representation of channel characteristics, encompassing delay and Doppler information. {It has been proved in \cite{AFDM_3} that  the positions of the non-zero entries in the channel matrix for each path do not overlap with each another,  $c_1$ should satisfy}\footnote{ { As pointed out in \cite{AFDM_3}, a larger value of $c_1$ generally results in increased channel estimation overhead. To  reduce the channel estimation overhead while achieving  a low  error rate performance, one can also choose  
  $c_1 =  \frac{{ 2{\alpha _{\max } + 1}}}{  2N_F\min_{i,i'}(\vert l_i  - {l_{i'}} \vert ) }$, i.e., $ k_{\varepsilon}=0$. }}
   {
  \begin{equation}
      c_1  \geq \frac{{2(\alpha _{\max } + k_{\varepsilon})}+1}{  2{N_F}    \underset{i,i'}{\min}(\vert l_i  - {l_{i'}} \vert )      }.
  \end{equation}}

\section{Performance Analysis}
\subsection{Spectral Efficiency}
The SE of the proposed AFDM-IM scheme can be formulated as
\begin{equation}
{\eta _{{\rm{AFDM - IM}}}} = \frac{{G(\lfloor{{\log }_2}{ { {\tbinom{N}{K}} } }\rfloor+ K{\log _2}(M))}}{{N_F}}.
\end{equation}
Additionally, the SE of the conventional AFDM is presented as
\begin{equation}
{\eta _{{\rm{AFDM}}}} = \log_2{(M)}.
\end{equation}

\vspace{-1em}
\subsection{Analysis of BER}
The received signal $\bf{y}$ in \eqref{AFDM_4} can be rewritten as
\begin{equation}
{\bf{y}} = {\sqrt{\rho}}\sum\limits_{i = 1}^P {{h_i}{{\bf{H}}_i}{\bf{x}} + {\bf{\tilde w}}}  = {\bf{\Phi }}({\bf{x}}){\bf{h}} + {\bf{\tilde w}},
\end{equation}
where ${\bf{h}} = [{h_1},{h_2},...,{h_P}]$ is a $P \times 1$ vector and ${\bf{\Phi }}({\bf{x}})$ is the $N_F \times P$ concatenated matrix
${\bf{\Phi }}({\bf{x}}) = \left[ {{{\bf{H}}_1}{\bf{x}}\left| {...\left| {{{\bf{H}}_P}{\bf{x}}} \right.} \right.} \right].$
The {conditional pair-wise error probability (PEP)} for the proposed AFDM-IM is
\begin{equation}
\begin{aligned}
P({\bf{x}} \to {\bf{\hat x}}\left| {\bf{h}} \right.) &= P\left( {\left\| {{\bf{y}} - {\sqrt{\rho}}{\bf{\Phi }}({\bf{\hat x}}){\bf{h}}} \right\|^2 < \left\| {{\bf{y}} - {\sqrt{\rho}}{\bf{\Phi }}({\bf{x}}){\bf{h}}} \right\|}^2 \right)\\
& = Q\left( {\sqrt {\frac{{\bf{\delta }}}{{2P{N_0}}}} } \right),
\end{aligned}
\end{equation}
where {$Q(x) = \frac{1}{\pi }\int_0^{{\pi  \mathord{\left/
 {\vphantom {\pi  2}} \right.
 \kern-\nulldelimiterspace} 2}} {\exp ( - \frac{{{x^2}}}{{2{{\sin }^2}\theta }}){\rm{d}}\theta }$} is the Gaussian Q-function and $
 {\bf{\delta }}= {\left\| {\sqrt{\rho}}{({\bf{\Phi }}({\bf{\hat x}}) - {\bf{\Phi }}({\bf{x}})){\bf{h}}} \right\|^2} = {{\bf{h}}^H}{\bf{\Psi h}}$ with ${\bf{\Psi }} = {\rho}{({\bf{\Phi }}({\bf{\hat x}}) - {\bf{\Phi }}({\bf{x}}))^H}({\bf{\Phi }}({\bf{\hat x}}) - {\bf{\Phi }}({\bf{x}}))$.

\begin{table}
\centering
\caption{{Delay-Doppler profile for different channels.}}
{\begin{tabular}{|c|c|c|c|}\hline
Parameter & 2-path & 3-path & 4-path \\\hline
Delay spread $l$ & (0,3) & (0,1,3) & (0,1,2,3)\\\hline
\makecell{normalized \\Doppler shift ${\varepsilon}$ }& (0.5, 0.8) & (0.2, 0.5, 0.7) & (0.2, 0.3, 0.5, 0.7)\\\hline
\end{tabular}}
\end{table}

The unconditional PEP  can be written as
{
\begin{equation}
\begin{aligned}
P\left( {{\bf{x}} \to {\bf{\hat x}}} \right) &= P({\bf{x}} \to {\bf{\hat x}}\left| {\bf{h}} \right.)\int_0^{ + \infty } {{p_{\bf{\delta }}}({\bf{\delta }})} {\rm{d}}{\bf{\delta }}\\
&=\frac{1}{\pi }\int_0^{{\pi  \mathord{\left/
 {\vphantom {\pi  2}} \right.
 \kern-\nulldelimiterspace} 2}} {\int_0^{ + \infty } {\exp } } \left( { - \frac{{\bf{\delta }} }{{4PN_0{{\sin }^2}\theta }}} \right){p_{\bf{\delta }} }({\bf{\delta }} ){\rm d}{\bf{\delta }} {\rm d}\theta \\
& = \frac{1}{\pi }\int_0^{\pi /2} {{M_{\bf{\delta }}}\left( { - \frac{1}{{4P{N_0}{{\sin }^2}\theta }}} \right)} {\rm{d}}\theta,
\end{aligned}
\end{equation}}
where ${M_{\bf{\delta }} }( s )$ is the moment generating function (MGF) of ${\bf{\delta }}$. Based on the MGF given in \cite{MGF}, we have
\begin{equation}
{M_{\bf{\delta }} }(s) = {\left| {{\bf{I}} - s{\bf{\Psi }}} \right|^{ - {1}}} = \prod\limits_{i = 1}^q {{{(1 - s{\lambda _i})}^{ - {1}}}},
\end{equation}
where $q$ is the number of the distinct non-zero eigenvalues ${\lambda _i}$ of the distance matrix ${\bf{\Psi}}$. As a result, the unconditional PEP is obtained as
\begin{equation}\label{product}
P\left( {{\bf{x}} \to {\bf{\hat x}}} \right) = \frac{1}{\pi }\int_0^{{\pi  \mathord{\left/
 {\vphantom {\pi  2}} \right.
 \kern-\nulldelimiterspace} 2}} {\prod\limits_{i = 1}^q {{{\left( {\frac{1}{{1 + \frac{{{\lambda _i}}}{{4PN_0{{\sin }^2}\theta }}}}} \right)}}} } {\rm d}\theta.
\end{equation}
The Chernoff bound of the PEP is obtained by substituting the integrand within \eqref{product} with its value at $\theta  = \pi /2$, resulting in
\begin{equation}
P\left( {{\bf{x}} \to {\bf{\hat x}}} \right) \le \prod\limits_{i = 1}^q {{{\left( {1 + \frac{{{\lambda _i}}}{{4{PN_0}}}} \right)}^{ - 1}}}.
\end{equation}
Therefore, the ABEP of the proposed AFDM-IM can be obtained by the asymptotically tight union upper bound as
\begin{equation}
{P_b} \le \frac{1}{{p{2^p}}}\sum\limits_{\bf{x}} {\sum\limits_{{\bf{\hat x}}} {P\left( {{\bf{x}} \to {\bf{\hat x}}} \right)} } e{\rm{(}}{\bf{x}} \to {\bf{\hat x}}{\rm{)}},
\end{equation}
where $e{\rm{(}}{\bf{x}} \to {\bf{\hat x}}{\rm{)}}$ is the Hamming distance of the corresponding pairwise error event.


\begin{figure}[t]
\centering
\includegraphics[width=0.4\textwidth]{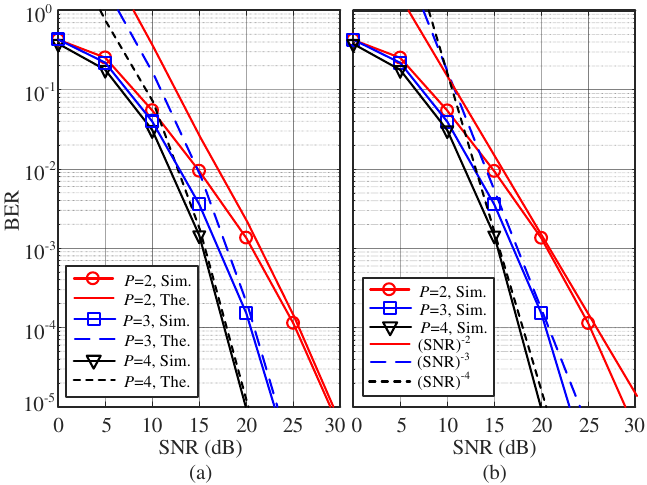}
\caption{Simulation and theoretical uncoded BER performance of the proposed AFDM-IM in different channels (i.e., path number $P=\{2,3,4\}$) using ML detector.}
\label{BER_Theory}
\end{figure}

\begin{figure}[t]
\centering
\includegraphics[width=0.4\textwidth]{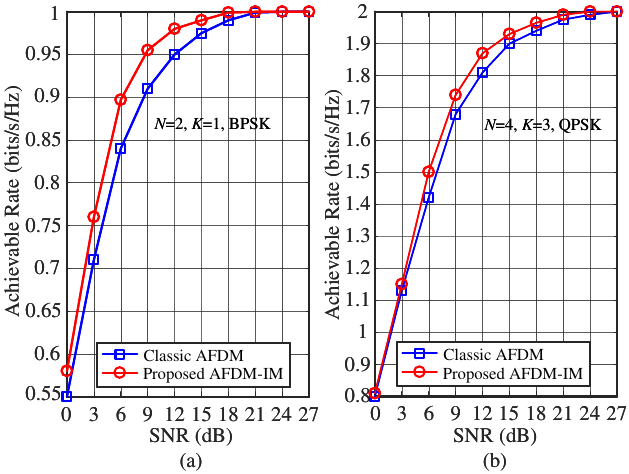}
\caption{{Achievable rate comparisons between the proposed AFDM-IM and the classic AFDM under channel uncoded scenario.}}
\label{AR}
\end{figure}

\begin{figure*}[h]
\centering
\vspace{-1em}
\includegraphics[width=1\textwidth]{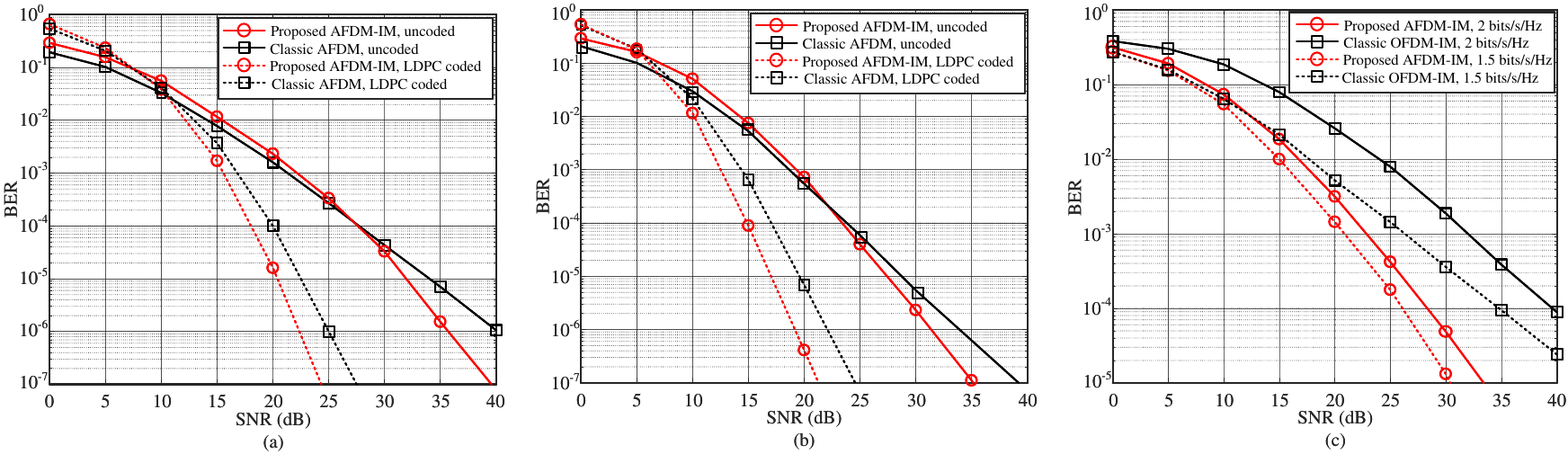}
\caption{{Uncoded/coded BER performance comparisons between the proposed ADFM-IM and the classic AFDM: (a) 2-path channel with 2 bits/s/Hz; (b) 4-path channel with 2 bits/s/Hz; (c) Uncoded  BER performance comparisons between the proposed AFDM-IM and the classic OFDM-IM under 2-path channel using MMSE detector.}}
\label{BER_UP}
\vspace{-1em}
\end{figure*}

\vspace{-1em}
\section{Simulation Results and Discussions}

{In this section, we present simulation results to examine the performance of the proposed AFDM-IM scheme. Here, the carrier frequency $f_c$ is 4 GHz and  chirp SC spacing $\Delta f$ is 2000 Hz.  We consider $ c_1  = \frac{{2(\alpha _{\max } + k_{\varepsilon})}+1}{2{N_F}}$ with $k_\epsilon =1$. In addition, we consider three different channel settings with different number of paths, i.e., $P=\{2,3,4\}$, namely 2-path, 3-path and 4-path channel, respectively, the delay-Doppler profiles for different channels are given in Table II}\footnote{{In the simulations, we define the signal-to-noise ratio (SNR) as $E_b/N_{0,T}$, where $E_b=\frac{E_T}{\eta N_F}$ is the average transmit power per bit. $N_{0,T}=\frac{N}{K}N_0$ is the noise variance in the time domain, which is related to the noise variance $N_0$ in the DAFT domain.}}.

Fig. \ref{BER_Theory} plots the analytical and simulated BER performance of the proposed AFDM-IM under different channel settings with $(N_F,N,K,M)=(32,8,1,2)$ using the ML detector. In particular, it can be observed from Fig. \ref{BER_Theory} (a) that the analytical BER curves approach the simulation results as the SNR increases. As expected, the proposed AFDM-IM can achieve full diversity order of the channel as shown in Fig. \ref{BER_Theory} (b), where asymptotic lines with slopes 2 $(\rm{SNR})^{-2}$, 3 $(\rm{SNR})^{-3}$ and 4 $(\rm{SNR})^{-4}$ are plotted.

{Fig. \ref{AR} compares the achievable rate of the proposed AFDM-IM and the classic AFDM at SEs of 1 and 2 bits/s/Hz. It can be observed from Fig. \ref{AR} that both the proposed AFDM-IM and the classic AFDM systems reach saturation in their achievable rates at their corresponding uncoded SEs at high SNR. In addition, the proposed AFDM-IM exhibits superior achievable rate performance compared to the classic AFDM. This is attributed to the fact that the bits associated with IM benefit from stronger protection compared to the ordinary modulation bits.}

{Figs. \ref{BER_UP} (a) and (b) compares the uncoded/coded BER performance of the proposed PR strategy aided AFDM-IM with the classic AFDM at the same SE $\eta=2$ bits/s/Hz. Here, the simulation parameter is set as $(N_F,N,K,M)=(64,4,3,4)$ for the proposed AFDM-IM, and $(N_F,M)=(64,4)$ for the classic AFDM. In addition, we use low density parity check (LDPC) code, where the code length is 256 and the code rate is $2/3$. It can be observed that the proposed AFDM-IM outperforms the classic AFDM in both channel coded/uncoded scenarios in the high SNR region. More precisely, in the high SNR region, the proposed scheme provides nearly 4 dB SNR gain over the classic AFDM at BER = $10^{-7}$ in the 4-path channel. The reasons are twofold: 1) the augmentation of the transmit power of the active SCs confers an enhanced capacity for detecting modulated symbols. 2) The inherent property of the detection for estimating the SC activity contributes to the overall improvement in the BER performance. In most cases only symbol detection error occurs in the high SNR region, while the estimation errors of SC activity dominate in the low SNR region but rarely encountered in the high SNR region. When comes to the PS strategy aided AFDM-IM, it saves $25 \%$ transmit power compared to the classic AFDM, since only a subset of SCs ($K=3$) are activated to transfer modulated symbols in the proposed scheme rather than all the SCs ($N=4$) are activated in the classic AFDM\footnote{{Note that the PS aided AFDM-IM is able to achieve the same BER performance as the PR aided AFDM-IM at the same normalized SNR region.}}.}


To further validate the advantages, {Fig. \ref{BER_UP} (c)} evaluates the uncoded BER performance of the proposed AFDM-IM and the conventional OFDM-IM with PS strategy under the same SE of $\eta=1.5/2$ bits/s/Hz. Specifically, we set $(N_F,N,K,M)=(64,4,3,4)$ to achieve the SE of 2 bits/s/Hz and $(N_F,N,K,M)=(128,2,1,2)$ to attain the SE of 1.5 bits/s/Hz for both the proposed AFDM-IM and the conventional OFDM-IM systems. Compared to the conventional OFDM-IM, the SNR gain for the proposed scheme with $\eta=2$ bits/s/Hz
is about 12 dB at the BER of $10^{-4}$, this is attributed to the inherent benefits of AFDM-IM, i.e., it can achieve full diversity gain over the time-varying channels. In other words, OFDM-IM cannot separate the paths, but the proposed scheme is capable of separating the paths by properly setting the DAFT parameters $c_1$ and $c_2$, thus exploiting the advantages of both the IM and AFDM techniques.

\section{Conclusion}
In this letter, we proposed a new multicarrier system, termed as AFDM-IM, which combines AFDM with IM to improve the system performance in terms of BER and EE. Then, two power allocation strategies were presented to enhance the BER and EE performance. Additionally, an ABEP upper bound was derived for the proposed AFDM-IM scheme and validated by the Monte Carlo simulations. Moreover, simulation results demonstrated the superiority of the proposed AFDM-IM scheme over the classic AFDM in terms of BER and EE. AFDM-IM constitutes a promising and viable modulation scheme for future wireless communication systems.

\end{document}